\documentstyle [12pt]{article}

\catcode`\@=11
\def\makepreprititle{\par
  \begingroup
  \def\thefootnote{\fnsymbol{footnote}}
  \def\
@makefnmark{\hbox
  to 0pt{$^{\@thefnmark}$\hss}}
  \if@twocolumn
  \twocolumn[\@makepreprititle]
  \else \newpage
  \global\@topnum\z@
  \@makepreprititle \fi\thispagestyle{empty}\@thanks
  \endgroup
  \setcounter{footnote}{0}
  \let\makepreprititle\relax
  \let\@makepreprititle\relax
  \gdef\@thanks{}\gdef\@author{}\gdef\@title{}
  \gdef\@preprintnumber{}\gdef\@preprintdate{}\gdef\subtitle{}
  \let\thanks\relax}
\def\preprintnumber#1{\gdef\@preprintnumber{#1}}
\def\preprintdate#1{\gdef\@preprintdate{#1}}
\def\subtitle#1{\gdef\@subtitle{#1}}
\def\@makepreprititle{\newpage
{\def\baselinestretch{1}
  \begin{flushright} \@preprintnumber \par
  \@preprintdate \end{flushright} } \par
  \begin{center}
\vskip 1.5em
  {\LARGE \@title \par} \vskip 2.5em
  {\Large \lineskip .5em
  \begin{tabular}[t]{c}\@author
  \end{tabular}\par}
  \vskip 1em {\large \@date} \end{center}
  \par
  \vfil}
\date{\sl Department of Physics, Tohoku University\\Sendai, 980 Japan}
\preprintdate{}
\preprintnumber{}
\subtitle{}
\def\abstract{\if@twocolumn
\section*{Abstract}
\else \normalsize
\begin{center}
{\bf Abstract\vspace{-.5em}\vspace{0pt}}
\end{center}
\quotation
\addtocounter{page}{-1}
\fi}
\def\endabstract{\if@twocolumn\else\endquotation\fi}
\catcode`\@=12
\renewcommand{\theequation}{\arabic{section}.\arabic{equation}}

\def\thebibliography#1{\section*
 {References					
 \markboth{REFERENCES}{REFERENCES}}\list
 {[\arabic{enumi}]}				
 {\settowidth\labelwidth{[#1]}\leftmargin\labelwidth
 \advance\leftmargin\labelsep
 \usecounter{enumi}}
 \def\newblock{\hskip .11em plus .33em minus -.07em}
 \sloppy \sfcode`\.=1000\relax}

\renewcommand{\theequation}{\thesection.\arabic{equation}}
\newcommand{\cleqn}{\setcounter{equation}{0}}
\newcommand{\clfn}{\setcounter{footnote}{0}}

\renewcommand{\thefootnote}{\fnsymbol{footnote}}
\newcommand \bra[1]{\left< {#1} \,\right\vert}
\newcommand \ket[1]{\left\vert\, {#1} \, \right>}

\newcommand{\bea}{\begin{eqnarray}}
\newcommand{\eea}{\end{eqnarray}}
\newcommand{\simgt}{\hbox{ \raise3pt\hbox to 0pt{$>$}\raise-3pt\hbox{$\sim$} }}
\newcommand{\simlt}{\hbox{ \raise3pt\hbox to 0pt{$<$}\raise-3pt\hbox{$\sim$} }}
\newcommand{\dal}{
\hbox{ \raise1pt\hbox to 0pt{$\sqcap$}\raise-1pt\hbox{$\sqcup$} }}
\newcommand \vc[1]{{\bf {#1}}}

\def\muhat{\hat{\mu}}
\def\nuhat{\hat{\nu}}
\def\to{\rightarrow}
\def\epem{\ifmmode{ e^{+}e^-} \else{$ e^{+}e^- $ } \fi}
\def\bw{\ifmmode{ bW^+ } \else{$ bW^+ $ } \fi}
\def\bwb{\ifmmode{ \bar{b}W^- } \else{$ \bar{b}W^- $ } \fi}
\def\ttbar{\ifmmode{t\bar{t}} \else{$t\bar{t}$} \fi}
\def\nrg{\ifmmode{\tilde{G}(\vc{p};E)} \else{$\tilde{G}(\vc{p};E)$} \fi}

\def\alpsmz{\alpha_{s}(m_Z)_{\overline{\bf MS}}}

\def \lsa {\rlap {\lower 3.5 pt \hbox {$\mathchar \sim$}} \raise 1
pt \hbox {$<$}}
\def \rsa {\rlap {\lower 3.5 pt \hbox {$\mathchar \sim$}} \raise 1
pt \hbox {$>$}}
\def\msbar {\ifmmode{\overline{\rm MS}}    \else{$\overline{\rm MS}$ }    \fi}

\def\alpsmz{\ifmmode{\alpha_s(m_Z)_{\overline{\rm MS}}}
              \else{$\alpha_s(m_Z)_{\overline{\rm MS}}$} \fi}
\def\lamfive{\ifmmode{\Lambda^{(5)}_{\overline{\rm MS}}}
               \else{$\Lambda^{(5)}_{\overline{\rm MS}}$} \fi}
\def\lamfour{\ifmmode{\Lambda^{(4)}_{\overline{\rm MS}}}
               \else{$\Lambda^{(4)}_{\overline{\rm MS}}$} \fi}
\def\lsa{\rlap{\lower 3.5 pt \hbox {$\mathchar \sim$}} \raise 1pt \hbox {$<$}}
\def\rsa{\rlap{\lower 3.5 pt \hbox {$\mathchar \sim$}} \raise 1pt \hbox {$>$}}

\begin{document}



\preprintnumber{TU--492}
\preprintdate{November, 1995}

\title{
New Method for Exact Calculation of Green Functions in
Scalar Field Theory
}

\author{Y.~Sumino}
\date{\sl  Department of Physics, Tohoku University, Sendai,
980-77 Japan}
\makepreprititle

\begin{abstract}
\normalsize

We present a
new method for calculating the Green functions for a lattice
scalar field theory in $D$ dimensions
with arbitrary potential $V(\phi)$.
The method for non-perturbative evaluation of Green functions for
$D \! = \! 1$ is generalized to higher dimensions.
We define ``hole functions'' $A^{(i)}~(i=0,1,2,\cdots,N \! -\! 1)$
from which one can construct $N$-point Green functions.
We derive characteristic equations of $A^{(i)}$
that form a {\it finite closed} set of coupled local equations.
It is shown that the Green functions constructed from the solutions to
the characteristic equations satisfy the Dyson-Schwinger equations.
To fix the boundary conditions of $A^{(i)}$,
a prescription is given for selecting the vacuum state at the
boundaries.

\end{abstract}
\vspace{1cm}

\vfil

\newpage

\normalsize

\section{Introduction}

In quantum field theory, various physical quantities are calculated from
the Green functions (correlation functions).
The principal
methods for calculating Green functions in a theory
for space-time dimensions $D \geq 3$ have been either to resort to
the perturbative expansion or to estimate their behavior in Euclidean
region by numerical simulation based on a lattice field theory.
It is desirable, however, that
the Green functions can be calculated other
than in perturbative expansion or without large computer calculation.

It is well-known that the Green functions obey
the Dyson-Schwinger equations, which constitute an
infinite hierarchy of equations.\cite{ds}
For a general interacting field theory, there is no finite
closed subset of these equations, which makes it difficult to solve
the equations exactly.
Meanwhile, for a scalar
field theory defined on a lattice
(with arbitrary potential $V(\phi)$),
we find there exist equations equivalent to the Dyson-Schwinger
equations that have finite closed subsets.
{}From the closed equations one may
calculate $N$-point Green functions, which can be used as a
new technique to evaluate Green functions.

The idea for finding the equations is
to generalize the method for non-perturbative evaluation of Green
functions in 1 dimension.
Consider
the 1-dimensional
(continuum) scalar field theory given by the action
\bea
S[\phi ] = \int_0^T dt \,
\left\{
\frac{1}{2} (\partial_t \phi)^2 - V(\phi)
\right\} ,
\label{1dimaction}
\eea
or equivalently,
the quantum mechanical system given by the Hamiltonian
\bea
\hat{H}_\phi = - \frac{1}{2} \frac{d^2}{d\phi^2} + V(\phi ) .
\label{1dhamil}
\eea
Suppose we want to calculate the one-point function defined by
\bea
G (t) = \bra{f} e^{-i\hat{H}(T-t)} \, \hat{\phi} \,
e^{-i\hat{H}t} \ket{i}
\eea
for some initial state $\ket{i}$ and final state $\ket{f}$.
For this purpose, first define a function
\bea
A(\varphi,\varphi';t) = \bra{f} e^{-i\hat{H}(T-t)} \ket{\varphi}
\bra{\varphi'} e^{-i\hat{H}t} \ket{i} ,
\label{heuristic1}
\eea
where
$\hat{\phi} \ket{\varphi^{(\prime )}} = \varphi^{(\prime )}
\ket{\varphi^{(\prime)}}$.
Once $A(\varphi,\varphi';t)$ is known,
the one-point function can be calculated as
\bea
G(t) = \int d\varphi \, \, \varphi \, A(\varphi,\varphi ;t) .
\label{gtfroma}
\eea
$A(\varphi , \varphi';t)$ satisfies a characteristic equation
\bea
i\frac{\partial}{\partial t} \, A(\varphi,\varphi';t) =
\left(
\hat{H}_{\varphi'} - \hat{H}_\varphi \right) A(\varphi,\varphi';t) .
\label{chareq}
\eea
Note that eq.(\ref{chareq}) is satisfied by
$A(\varphi , \varphi';t)$ corresponding to any initial and final states,
$\ket{i}$ and $\ket{f}$.
As we will see later, there is a prescription to select a particular
solution to eq.(\ref{chareq}) for which both $\ket{i}$ and $\ket{f}$ are
the ground state $\ket{0}$
of the Hamiltonian (\ref{1dhamil}).
So, in fact we may evaluate
the one-point function $\bra{0} \hat{\phi} \ket{0}$
by solving eq.(\ref{chareq}).

The simplest Dyson-Schwinger equation related to $G(t)$ is given by
\bea
\frac{d^2}{dt^2} \, G(t) =
- \int d\varphi \, V'(\varphi ) \, A(\varphi,\varphi ;t) ,
\label{ehrenfest}
\eea
which is just the Ehrenfest formula when $\ket{i}=\ket{f}$.
Using fundamental techniques of quantum mechanics, it is easy to prove
eq.(\ref{ehrenfest}) for any $A(\varphi ,\varphi' ;t)$ satisfying
eq.(\ref{chareq}) and for $G(t)$ constructed via eq.(\ref{gtfroma}).

It is eqs.(\ref{heuristic1}) and (\ref{chareq}) that we are going to
generalize for scalar field theory regularized on a
lattice in higher dimensions.
We will define a
``hole function'' $A_n$ associated with a lattice site
$n$, and will see that it obeys local equations that characterize
local property of the field theory in a non-trivial way.

Most of the discussion given in this paper holds in parallel both for
Minkowski and Euclidean space-time metrics.
For notational convenience, we will adopt the Euclidean theory in the
following, and we list Minkowski versions of some important equations
in Appendix A.

In Section 2 we review the Dyson-Schwinger equations of
a lattice scalar
field theory.
In Section 3 we will define a hole function, from which Green functions
can be calculated, and derive local equations satisfied by the hole
function.
We show that these local equations are equivalent to the Dyson-Schwinger
equations in Section 4.
Then we discuss in Section 5
how to extract Green functions satisfying a
specific boundary conditions, namely those given by the vacuum
expectation values of field operator products.
Some basic properties of the local equations are briefly summarized in
Section 6.
Section 7 is devoted to summary and discussion.

We list some equations for Minkowski space-time in Appendix A.
Proof of formulas used in Section 4 is given in
Appendix B.

\section{Dyson-Schwinger Equations}
\cleqn

We start\footnote{
We will not pay attention to the boundary conditions of the field
configurations until Section 5.
For definiteness, one may assume periodic boundary conditions for
discussion in Sections 2 - 4.
}
by reviewing the Dyson-Schwinger equations for a
Euclidean scalar
field theory defined on a $D$-dimensional lattice $\Lambda^D$
with a lattice spacing $\epsilon$.
The $N$-point Green function for the lattice scalar field theory is
given by
\bea
\left< \phi_{i_1} \cdots \phi_{i_N} \right>
= \int \prod_{l \in \Lambda^D} d \phi_l ~
\phi_{i_1} \cdots \phi_{i_N} \, e^{-S[\phi ]}
\biggl/
\int \prod_{l \in \Lambda^D} d \phi_l ~ e^{-S[\phi ]}
,
\label{defgreen}
\eea
where the
action is defined as the discretized version of
eq.(\ref{1dimaction}) for $D$-dimensional Euclidean space-time:
\bea
S[\phi ] =
\sum_{l \in \Lambda^D} \epsilon^D
\biggl\{ \,
\sum_{\muhat =1}^D \frac{1}{2}
\biggl( \frac{\phi_{l+\muhat} - \phi_l}{\epsilon} \biggl)^2
+ V(\phi_l ) \, \biggl\} .
\label{defaction}
\eea
Here and hereafter, $\muhat$ denotes a unit vector
in one of $2D$ directions
$(\pm \hat{x}_1, \ldots, \pm \hat{x}_D )$.
For notational simplicity, the following rules for the sum of directions
are understood throughout the paper.
\bea
\begin{array}{ccl}
{\displaystyle \rule[-9mm]{0mm}{1mm}
\sum_{\muhat =1}^{D} }&:&
\mbox{sum over}~ \muhat =  +\hat{x}_1, \ldots, +\hat{x}_D , \\
{\displaystyle
\sum_{\muhat =1}^{2D} }&:&
\mbox{sum over}~ \muhat = \pm \hat{x}_1, \ldots, \pm \hat{x}_D . \\
\end{array}
\eea

The expression for the Green function
has a simple form of taking the expectation value of
field product $\phi_{i_1} \cdots \phi_{i_N}$ with the weight factor
$e^{-S[\phi ]}$.
In the following,
we derive the relationship among the Green functions starting
from defining equations of this weight factor
$w \equiv e^{-S[\phi ]}$:
\bea
\frac{\partial}{\partial \phi_n} \, \left( e^{S[\phi ]} \, w \right) = 0
{}~~~~~
\mbox{for }\forall n ,
\label{defw1}
\eea
or,
\bea
\left[
\frac{1}{\epsilon^D}
\frac{\partial}{\partial \phi_n} - \dal_n \phi_n
+ V'( \phi_n )
\right] \, w = 0 ,
\label{defw2}
\eea
where
\bea
\dal_n X_n = \sum_{\muhat =1}^D
\left(
\frac{X_{n+\muhat} - 2 X_n + X_{n-\muhat}}{\epsilon^2}
\right)
=
\sum_{\muhat =1}^{2D}
\left(
\frac{X_{n+\muhat} - X_n}{\epsilon^2}
\right) .
\eea
{}From the first equations (\ref{defw1}),
it is clear that the equations define $w$ uniquely up to
a physically unimportant ($\phi$-independent) coefficient.

Next, we Fourier transform the defining equations
(\ref{defw2}) with respect
to all $\phi_l$'s as
\bea
0 &=&
\int \prod_{l} d \phi_l ~
\exp \biggl( i \epsilon^D \sum_{l} J_l \phi_l \biggl) \,
\left[
\frac{1}{\epsilon^D}
\frac{\partial}{\partial \phi_n} - \dal_n \phi_n
+ V'( \phi_n )
\right] \, e^{-S[\phi ]}
\\
&=&
\int \prod_{l} d\phi_l ~
\left[
-i J_n - \dal_n \phi_n + V'( \phi_n )
\right] \, e^{-S[\phi ] + i \epsilon^D \sum J_l \phi_l } .
\eea
Replacing $\phi_n$ by $\partial /\partial J_n$, we obtain the
following coupled partial differential equations for the partition
function, that is, Dyson-Schwinger equations:
\bea
&&
\left\{
\dal_n \biggl( \frac{1}{i\epsilon^D}
\frac{\partial}{\partial J_n} \biggl)
- V' \biggl( \frac{1}{i\epsilon^D}
\frac{\partial}{\partial J_n} \biggl) + i J_n
\right\}
Z[J] = 0 ,
\label{sdeq1}
\\
&&
Z[J] =
\int \prod_{l} d\phi_l ~
e^{-S[\phi ] + i \epsilon^D \sum J_l \phi_l } .
\label{defpfn}
\eea
Perhaps the
physical meaning of the equations is most transparent when
we regard them as the expectation values of the equations
of motion in the
presence of source $J$:
\bea
\left<
\dal_n \phi_n - V'( \phi_n ) + i J_n
\right>_J = 0 .
\eea
Expanding the partition function in Taylor series in $J$ as
\bea
Z[J] = Z[0] \times
\sum_{N=0}^\infty \frac{(i\epsilon)^D}{N!} \,
\sum_{i_1 \in \Lambda^D} \cdots \sum_{i_N \in \Lambda^D} \,
J_{i_1} \cdots J_{i_N} \,
\left< \phi_{i_1} \cdots \phi_{i_N} \right> ,
\eea
and substituting to eq.(\ref{sdeq1}), we obtain
coupled equations among the Green functions:
\bea
\dal_n \left< \phi_n \, \phi_{i_1} \cdots \phi_{i_N} \right>
- \left< V'(\phi_n) \, \phi_{i_1} \cdots \phi_{i_N} \right>
= -
\sum_{k=1}^N \, \frac{1}{\epsilon^D} \, \delta_{n,i_k}
\left< \phi_{i_1} \cdots \right. \! \!
{\hbox{ \raise-5pt\hbox to 0pt{$\frown$}\raise-10pt\hbox{$\phi_{i_k}$} }}
\! \! \left. \cdots \phi_{i_N} \right> .
\label{sdeq2}
\eea

Since eqs.(\ref{sdeq1}) are nothing but Fourier transform
of the defining equations for
the weight factor $e^{-S}$,
Green functions obtained by solving
the Dyson-Schwinger equations (\ref{sdeq2})
are (almost) equivalent to the Green functions
defined by the integral eq.(\ref{defgreen}).\footnote{
Note that eq.(\ref{sdeq1}) holds for any contour
in the complex
$\phi$-plane of $\phi$-integral
in eq.(\ref{defpfn}).
Therefore, there is
additional degree of
freedom for solutions to the Dyson-Schwinger
equations compared with the original definition (\ref{defgreen}).
The degree of freedom
corresponds to the number of independent contours in the complex
$\phi$-plane for each $\phi_l$-integration.
}
In this sense, the Dyson-Schwinger equations contain
full information on
the lattice field theory.

Another important feature of the Dyson-Schwinger equations is that if
one tries to solve the equations to obtain the $N$-point
Green function for some $N$, one in fact needs to solve all
the coupled equations.
Namely, the coupled equations never close with some
finite subset of the equations, so one
needs to deal with infinite-dimensional coupled equations.

\section{Hole Function and Local Equations}
\cleqn

We consider the lattice scalar field theory as defined
in the previous section.
For later convenience, we make a slight change in the integral
measure of
the partition function
\bea
Z[J] = \int \prod_{l \in \Lambda^D} [d \phi_l] ~
\exp \biggl[
- S[\phi] + i\epsilon^D \sum_{l} J_l \phi_l
\biggl] ,
\label{pfn2}
\eea
where $[d\phi_l] = d\phi_l/\sqrt{2\pi\epsilon}$.
(This change does not alter the Green functions (\ref{defgreen}).)
The action is given by eq.(\ref{defaction}).

Let us take one site $n \in \Lambda^D$, and define the
``hole function of $n$'' as a function of $2D$ link variables
$u_1, \ldots , u_{2D}$ surrounding the site $n$ (Fig.1) by
\bea
&&
A_n (u_1, \cdots , u_{2D}; J)
\nonumber
\\
&& \equiv
\int \prod_{l \neq n} [ d\phi_l ] \,
\exp \biggl[
- S_n'[\phi]
+ i\epsilon^D \sum_{l \neq n}J_l \phi_l
- \epsilon^D \sum_{\muhat = 1}^{2D} \frac{1}{2}
\biggl( \frac{\phi_{n+\muhat}-u_{\muhat}}{\epsilon} \biggl)^2
\biggl] ,
\label{defholefn}
\eea
where $S_n'[\phi]$ denotes the part of the
action $S[\phi]$ that remains after subtraction of terms
depending on $\phi_n$:
\bea
S_n'[\phi] = S[\phi] - \epsilon^D
\biggl\{ \,
\sum_{\muhat =1}^{2D} \frac{1}{2}
\biggl( \frac{\phi_{n+\muhat} - \phi_n}{\epsilon} \biggl)^2
+ V(\phi_l ) \, \biggl\} .
\eea
As compared to $Z[J]$, not only we leave $\phi_n$ unintegrated but also
we subtracted all $\phi_n$-dependent terms in the exponent
in eq.(\ref{defholefn}) first and then re-added the
kinetic term after
replacing $\phi_n$ by the link variables $u_1, \ldots , u_{2D}$.
Using this hole function, expectation value of a local operator
can be expressed as
\bea
\left< \phi_n^{\, k} \right>_J =
\frac{
\rule[-2mm]{0mm}{6mm}
\int [du] \, u^k \,
A_n (u, \cdots, u;J) \,
e^{-\epsilon^D V(u ) + i\epsilon^D J_n u }
}{
\rule[0mm]{0mm}{5mm}
\int [du] \,
A_n (u, \cdots, u;J) \,
e^{-\epsilon^D V(u ) + i\epsilon^D J_n u }
} .
\eea
Also, we can
construct the Green functions
from the hole function as
\bea
\left< \phi_n \, \phi_{i_1} \cdots \phi_{i_N} \right>
=
\frac{
{\displaystyle \int}
[du] \, \, u \,
\left[ {\textstyle
\frac{1}{i\epsilon^D}
\frac{\partial}{\partial J_{i_1}} \ldots
\frac{1}{i\epsilon^D}
\frac{\partial}{\partial J_{i_N}}
\, A_n (u, \cdots , u;J )
} \right]_{J=0}
e^{-\epsilon^D V(u)}
}{
{\displaystyle \int}
[du] \,  \left[ A_n (u, \cdots , u;J)
\right]_{J=0}
e^{-\epsilon^D V(u)}
}
{}.
\label{greenfnfroma}
\eea
Note that
from the definition (\ref{defholefn})
the hole function has a property
\bea
A_n (u_1, \cdots , u_{2D}; J)
= \, \mbox{independent of}~J_n ,
\label{condfora}
\eea
although it depends on $J_l$ for $l \neq n$.

We are going to derive a local equation satisfied by
the above hole function $A_n$.
For this purpose, we define a function associated with
the two adjacent sites
$n$ and $n+\muhat$ as follows.
(See Fig.2.)
\bea
F_{n,\muhat} (u_1,
{\hbox{ \raise-5pt\hbox to 0pt{$\frown$}\raise-14pt
\hbox to -3pt{$u_{\muhat}$} }}
\! \! \cdots, u_{2D} \, ; \tilde{u}_1,
{\hbox{ \raise-5pt\hbox to 0pt{$\frown$}\raise-14pt
\hbox to -3pt{$\tilde{u}_{-\muhat}$} }}
\! \! \cdots, \tilde{u}_{2D} \, ;J) ~~
\rule[0mm]{0mm}{6mm}
= \int \prod_{l \neq n, n+\muhat} \! \! [ d\phi_l ] \,
\exp \biggl[
- S_{n,n+\muhat}''[\phi]
+ i\epsilon^D \! \! \sum_{l \neq n, n+\muhat} \! \! J_l \phi_l
\nonumber
\\
- \epsilon^D \sum_{\nuhat \neq \muhat}^{2D-1} \frac{1}{2}
\biggl( \frac{\phi_{n+\nuhat}-u_{\nuhat}}{\epsilon} \biggl)^2
- \epsilon^D \sum_{\nuhat \neq -\muhat}^{2D-1} \frac{1}{2}
\biggl(
\frac{\phi_{n+\muhat + \nuhat}-\tilde{u}_{\nuhat}}{\epsilon}
\biggl)^2
\biggl] .
\eea
Here, $S_{n,n+\muhat}''[\phi] $
denotes the part of the
action $S[\phi]$ that remains after subtraction of terms
depending on $\phi_n$ and $\phi_{n+\muhat}$.
The function $F_{n,\muhat}$ depends on the variables on the links
surrounding the sites $n$ and $n+\muhat$ but
the one connecting $n$ and $n+\muhat$.
This function is defined such
that both $A_n$ and $A_{n+\muhat}$ can be
constructed from it.

The hole function $A_n$ is obtained from
$F_{n,\muhat}$ by integrating over the field variable on the site
${n+\muhat}$:
\bea
A_n(u_1,\cdots,u_{2D};J) &=&
\int [d\tilde{u}] \,
\exp \biggl[
-\epsilon^D \biggl\{
\frac{1}{2} \biggl( \frac{u_{\muhat}-\tilde{u}}{\epsilon}
\biggl)^2 + V(\tilde{u})
\biggl\}
+ i\epsilon^D J_{n+\muhat} \tilde{u} \biggl]
\nonumber
\\ &&~~~~~~~~~~~~~~~~~~
\times
\rule[0mm]{0mm}{6mm}
F_{n,\muhat} (u_1,
{\hbox{ \raise-5pt\hbox to 0pt{$\frown$}\raise-14pt
\hbox to -3pt{$u_{\muhat}$} }}
\! \! \cdots, u_{2D} \, ; \tilde{u},
\cdots, \tilde{u} \, ;J)
\label{relaandf1}
\\
&=&
\rule[0mm]{0mm}{8mm}
\exp \biggl[ \,
\frac{1}{2} \epsilon^{2-D}
\frac{\partial^2}{\partial u_{\muhat}^2}
\, \biggl] \,
\exp \biggl[
- \epsilon^D V(u_{\muhat}) + i\epsilon^D J_{n+\muhat} u_{\muhat}
\biggl]
\nonumber
\\ && ~~~~~~~~~~~~~~~~
\rule[0mm]{0mm}{6mm}
\times
F_{n,\muhat} (u_1,
{\hbox{ \raise-5pt\hbox to 0pt{$\frown$}\raise-14pt
\hbox to -3pt{$u_{\muhat}$} }}
\! \! \cdots, u_{2D} \, ; u_{\muhat},
\cdots, u_{\muhat} \, ;J) ,
\label{relaandf2}
\eea
where in the second line we used the identity\footnote{
To derive the integral form (left-hand side) from the differential form
(right-hand side), substitute
\bea
f(x) = \int dy \, \delta (x-y) \, f(y)
= \int \frac{dp \, dy}{2\pi} \, e^{ip(x-y)} \, f(y)
\nonumber
\eea
and integrate over $p$ after replacing $d/dx$ by $ip$.

Also, one may show a similar identity for the
inverse transformation:
\bea
\int^\infty_{-\infty} \frac{dy}{\sqrt{2\pi a}} \,
e^{-(y+ix)^2/2a} \, f(iy)
= e^{-\frac{1}{2}a \frac{d^2}{dx^2}} \, f(x)
{}~~~~~
(a>0) .
\nonumber
\eea
}
\bea
\int \frac{dy}{\sqrt{2\pi a}} \,
e^{-(x-y)^2/2a} \, f(y) =
e^{\frac{1}{2}a\frac{d^2}{dx^2}} \, f(x)
{}~~~~~
(a>0) .
\label{difint}
\eea
Then we can easily invert eq.(\ref{relaandf2}) to find
\bea
&&
F_{n,\muhat} (u_1,
{\hbox{ \raise-5pt\hbox to 0pt{$\frown$}\raise-14pt
\hbox to -3pt{$u_{\muhat}$} }}
\! \! \cdots, u_{2D} \, ; u_{\muhat},
\cdots, u_{\muhat} \, ;J)
\nonumber
\\
&&
= \exp \biggl[ \,
\epsilon^D V(u_{\muhat}) - i\epsilon^D J_{n+\muhat} u_{\muhat}
\, \biggl]
\,
\exp \biggl[ \,
- \frac{1}{2} \epsilon^{2-D}
\frac{\partial^2}{\partial u_{\muhat}^2}
\, \biggl] \,
A_n(u_1,\cdots,u_{2D};J) .
\label{relaandf3}
\eea
On the other hand, one may express the hole function $A_{n+\muhat}$ in
terms of $F_{n,\muhat}$:
\bea
A_{n+\muhat} (\tilde{u}_1, \cdots, \tilde{u}_{2D};J )
&=&
\int [du] \,
\exp \biggl[
-\epsilon^D \biggl\{
\frac{1}{2} \biggl( \frac{\tilde{u}_{-\muhat}-u}{\epsilon}
\biggl)^2 + V(u)
\biggl\}
+ i\epsilon^D J_{n} u \biggl]
\nonumber
\\ &&~~~~~~~~~~~~~~~~~~
\times
\rule[0mm]{0mm}{6mm}
F_{n,\muhat} (u, \cdots, u \, ;
\tilde{u}_1,
{\hbox{ \raise-5pt\hbox to 0pt{$\frown$}\raise-14pt
\hbox to -3pt{$\tilde{u}_{-\muhat}$} }}
\! \! \cdots, \tilde{u}_{2D} \, ;J) .
\label{relaandf4}
\eea
Using representation by differential operator as before, we obtain
\bea
&&
F_{n,\muhat} (\tilde{u}_{-\muhat},
\cdots, \tilde{u}_{-\muhat} \, ;
\tilde{u}_1,
{\hbox{ \raise-5pt\hbox to 0pt{$\frown$}\raise-14pt
\hbox to -3pt{$\tilde{u}_{-\muhat}$} }}
\! \! \cdots, \tilde{u}_{2D} \, ; J)
\nonumber
\\
&&
= \exp \biggl[ \,
\epsilon^D V(\tilde{u}_{-\muhat}) - i\epsilon^D J_{n} \tilde{u}_{-\muhat}
\, \biggl]
\,
\exp \biggl[ \,
- \frac{1}{2} \epsilon^{2-D}
\frac{\partial^2}{\partial \tilde{u}_{-\muhat}^2}
\, \biggl] \,
A_{n+\muhat}(\tilde{u}_1,\cdots,\tilde{u}_{2D};J) .
\label{relaandf5}
\eea
Comparing eqs.(\ref{relaandf3}) and (\ref{relaandf5}), we find
\bea
&&
F_{n,\muhat} (u,\cdots,u \, ; \tilde{u}, \cdots, \tilde{u} \, ;J)
\nonumber
\\
&&
= \exp \biggl[ \,
\epsilon^D V(\tilde{u}) - i\epsilon^D J_{n+\muhat} \tilde{u}
\, \biggl]
\,
\exp \biggl[ \,
- \frac{1}{2} \epsilon^{2-D}
\frac{\partial^2}{\partial \tilde{u}^2}
\, \biggl] \,
A_n(u,\cdots,u,
{\hbox to 0pt{$\tilde{u}$}\raise-7pt
\hbox to 0pt{$\, {\scriptstyle \uparrow}$}
\raise-17pt \hbox to 9pt{$\, {\scriptstyle \muhat}$}},
u, \cdots, u \, ;J)
\nonumber
\\
&&
= \exp \biggl[ \,
\epsilon^D V(u) - i\epsilon^D J_{n} u
\, \biggl]
\,
\exp \biggl[ \,
- \frac{1}{2} \epsilon^{2-D}
\frac{\partial^2}{\partial u^2}
\, \biggl] \,
A_{n+\muhat}(\tilde{u},\cdots,\tilde{u},
{\hbox to 0pt{$u$}\raise-7pt
\hbox to -2pt{$\, {\scriptstyle \uparrow}$}
\raise-17pt \hbox to 9pt{${\scriptstyle -\muhat}$}},
\tilde{u}, \cdots, \tilde{u} \, ;J) .
\nonumber
\\
\eea
Or, equivalently,
\bea
&&
\exp \biggl[ \,
\frac{1}{2} \epsilon^{2-D}
\frac{\partial^2}{\partial u^2}
\, \biggl] \,
\exp \biggl[ \,
- \epsilon^D V(u) + i\epsilon^D J_{n} u
\, \biggl]
\,
A_n(u,\cdots,u,
{\hbox to 0pt{$\tilde{u}$}\raise-7pt
\hbox to 0pt{$\, {\scriptstyle \uparrow}$}
\raise-17pt \hbox to 9pt{$\, {\scriptstyle \muhat}$}},
u, \cdots, u \, ;J)
\nonumber
\\
&&
=
\exp \biggl[ \,
\frac{1}{2} \epsilon^{2-D}
\frac{\partial^2}{\partial \tilde{u}^2}
\, \biggl] \,
\exp \biggl[ \,
- \epsilon^D V(\tilde{u}) + i\epsilon^D J_{n+\muhat} \tilde{u}
\, \biggl]
\,
A_{n+\muhat}(\tilde{u},\cdots,\tilde{u},
{\hbox to 0pt{$u$}\raise-7pt
\hbox to -2pt{$\, {\scriptstyle \uparrow}$}
\raise-17pt \hbox to 9pt{${\scriptstyle -\muhat}$}},
\tilde{u}, \cdots, \tilde{u} \, ;J) .
\nonumber
\\
\label{localeq}
\eea
These are the local equations
satisfied by the hole function we set out to
derive.
Using eq.(\ref{difint}) we obtain
coupled linear integral equations.

We will show in the next section that the local equations
(\ref{localeq}) together with the condition (\ref{condfora}) are
equivalent to the Dyson-Schwinger equations.
In marked contrast to the
Dyson-Schwinger equations, however,
we obtain sets of {\it closed} equations among the hole functions
when we expand eqs.(\ref{localeq})
in Taylor series in $J$.
For example,
if we define
\bea
A_n^{(0)} & \equiv & A_n \biggl|_{J=0} ,
\rule[-4mm]{0mm}{6mm}
\label{0thholefn}
\\
A_{n;k}^{(1)} & \equiv &
\frac{1}{i\epsilon^D} \,
\frac{\partial}{\partial J_k} \, A_n \biggl|_{J=0} ,
\label{1stholefn}
\eea
then the zeroth order and the first order equations, respectively,
become as
follows.
\bea
&&
\exp \biggl[ \,
\frac{1}{2} \epsilon^{2-D}
\frac{\partial^2}{\partial u^2}
\, \biggl] \,
\exp \biggl[ \,
- \epsilon^D V(u)
\, \biggl]
\,
A_n^{(0)}(u,\cdots,u,
{\hbox to 0pt{$\tilde{u}$}\raise-7pt
\hbox to 0pt{$\, {\scriptstyle \uparrow}$}
\raise-17pt \hbox to 9pt{$\, {\scriptstyle \muhat}$}},
u, \cdots, u )
\nonumber
\\
&&=
\exp \biggl[ \,
\frac{1}{2} \epsilon^{2-D}
\frac{\partial^2}{\partial \tilde{u}^2}
\, \biggl] \,
\exp \biggl[ \,
- \epsilon^D V(\tilde{u})
\, \biggl]
\,
A_{n+\muhat}^{(0)}(\tilde{u},\cdots,\tilde{u},
{\hbox to 0pt{$u$}\raise-7pt
\hbox to -2pt{$\, {\scriptstyle \uparrow}$}
\raise-17pt \hbox to 9pt{${\scriptstyle -\muhat}$}},
\tilde{u}, \cdots, \tilde{u} )
\label{zeroth}
\eea
and
\bea
&&
\exp \biggl[ \,
\frac{1}{2} \epsilon^{2-D}
\frac{\partial^2}{\partial u^2}
\, \biggl] \,
\exp \biggl[ \,
- \epsilon^D V(u)
\, \biggl]
\,
A_{n;k}^{(1)}(u,\cdots,u,
{\hbox to 0pt{$\tilde{u}$}\raise-7pt
\hbox to 0pt{$\, {\scriptstyle \uparrow}$}
\raise-17pt \hbox to 9pt{$\, {\scriptstyle \muhat}$}},
u, \cdots, u )
\nonumber
\\
&&
-
\exp \biggl[ \,
\frac{1}{2} \epsilon^{2-D}
\frac{\partial^2}{\partial \tilde{u}^2}
\, \biggl] \,
\exp \biggl[ \,
- \epsilon^D V(\tilde{u})
\, \biggl]
\,
A_{n+\muhat ;k}^{(1)}(\tilde{u},\cdots,\tilde{u},
{\hbox to 0pt{$u$}\raise-7pt
\hbox to -2pt{$\, {\scriptstyle \uparrow}$}
\raise-17pt \hbox to 9pt{${\scriptstyle -\muhat}$}},
\tilde{u}, \cdots, \tilde{u} )
\nonumber \\
&=&
\delta_{n+\muhat,k} \,
\exp \biggl[ \,
\frac{1}{2} \epsilon^{2-D}
\frac{\partial^2}{\partial \tilde{u}^2}
\, \biggl] \,
\exp \biggl[ \,
- \epsilon^D V(\tilde{u})
\, \biggl]
\, \tilde{u} \,
A_{n+\muhat}^{(0)}(\tilde{u},\cdots,\tilde{u},
{\hbox to 0pt{$u$}\raise-7pt
\hbox to -2pt{$\, {\scriptstyle \uparrow}$}
\raise-17pt \hbox to 9pt{${\scriptstyle -\muhat}$}},
\tilde{u}, \cdots, \tilde{u} )
\nonumber \\
&& -
\delta_{n,k} \,
\exp \biggl[ \,
\frac{1}{2} \epsilon^{2-D}
\frac{\partial^2}{\partial u^2}
\, \biggl] \,
\exp \biggl[ \,
- \epsilon^D V(u)
\, \biggl]
\, u \,
A_n^{(0)}(u,\cdots,u,
{\hbox to 0pt{$\tilde{u}$}\raise-7pt
\hbox to 0pt{$\, {\scriptstyle \uparrow}$}
\raise-17pt \hbox to 9pt{$\, {\scriptstyle \muhat}$}},
u, \cdots, u )  .
\label{first}
\eea
Also, a condition follows from eq.(\ref{condfora}):
\bea
A_{n;n}^{(1)} (u_1,\cdots,u_{2D}) = 0 .
\label{cond}
\eea
So, if we solve eqs.(\ref{zeroth})-(\ref{cond}),
we can calculate one-point and
two-point Green functions, respectively, using eq.(\ref{greenfnfroma}).

\section{Equivalence to the Dyson-Schwinger Equations}
\cleqn

We have seen in the previous section
that the hole function defined in eq.(\ref{defholefn})
obey the local equations (\ref{localeq}).
Conversely, one may define a hole function to be the solution to the
local equations (\ref{localeq}) satisfying the condition
(\ref{condfora}).
Now we are going to show that
the hole function defined in this way generates the partition function
that obey the Dyson-Schwinger
equations.

Let us begin by showing that ${\cal Z}_n$ defined as
\bea
{\cal Z}_n = \int [du] \,
\exp \left[ -\epsilon^D V(u) + i\epsilon^D J_n u \right]
\,
A_n(u, \cdots, u; J)
\eea
is independent of $n$.
This property would have been
trivial if we had adopted the definition (\ref{defholefn}) since
${\cal Z}_n$ is just the partition function $Z[J]$.
Using the local equations (\ref{localeq}), we see
\bea
{\cal Z}_{n+\muhat} &=& \int [du] \,
\exp \left[ -\epsilon^D V(u) + i\epsilon^D J_{n+\muhat} u \right]
\,
A_{n+\muhat}(u, \cdots, u; J)
\nonumber
\\
&=&
\int [du] \,
\biggl\{
\exp \left[ -\epsilon^D V(\tilde{u})
+ i\epsilon^D J_{n+\muhat} \tilde{u} \right] \,
A_{n+\muhat}(\tilde{u},\cdots,\tilde{u},
{\hbox to 0pt{$u$}\raise-7pt
\hbox to -2pt{$\, {\scriptstyle \uparrow}$}
\raise-17pt \hbox to 9pt{${\scriptstyle -\muhat}$}},
\tilde{u}, \cdots, \tilde{u} \, ;J)
\biggl\}_{\displaystyle \tilde{u} \to u}
\nonumber
\\
&=&
\int [du] \,
\biggl\{
\exp \biggl[ \,
-\frac{1}{2} \epsilon^{2-D}
\frac{\partial^2}{\partial \tilde{u}^2}
+\frac{1}{2} \epsilon^{2-D}
\frac{\partial^2}{\partial u^2}
\, \biggl]
\nonumber
\\
&& ~~~~~~~~~~
\times
\exp \biggl[ \,
- \epsilon^D V(u) + i\epsilon^D J_{n} u
\, \biggl]
\,
A_n(u,\cdots,u,
{\hbox to 0pt{$\tilde{u}$}\raise-7pt
\hbox to 0pt{$\, {\scriptstyle \uparrow}$}
\raise-17pt \hbox to 9pt{$\, {\scriptstyle \muhat}$}},
u, \cdots, u \, ;J)
\biggl\}_{\displaystyle \tilde{u} \to u}
\nonumber
\\
&=&
\int [du] \,
\exp \left[ -\epsilon^D V(u) + i\epsilon^D J_n u \right]
\,
A_n(u, \cdots, u; J)
=
{\cal Z}_n ,
\label{relzn}
\eea
where in the last line we used the identity
\bea
&&
\int dx \,
\biggl\{
\exp \biggl[ \,
-\frac{1}{2} \epsilon^{2-D}
\frac{\partial^2}{\partial y^2}
+\frac{1}{2} \epsilon^{2-D}
\frac{\partial^2}{\partial x^2}
\, \biggl] \,
f(x,y)
\biggl\}_{\displaystyle y \to x}
=
\int dx \, f(x,x) .
\label{id1}
\eea
(See Appendix B for proof.)
Thus, ${\cal Z}_n$ is independent of $n$,
so we will denote ${\cal Z}_n = Z[J]$ in the following.

We can derive yet another local property of $Z[J]$
in a similar manner.
Again using eq.(\ref{localeq}), we find
\bea
&&
\int [du] \, \, u \,
\exp \left[ -\epsilon^D V(u) + i\epsilon^D J_{n+\muhat} u \right]
\,
A_{n+\muhat}(u, \cdots, u; J)
\nonumber
\\
&&=
\int [du] \, \, u \,
\biggl\{
\exp \left[ -\epsilon^D V(\tilde{u})
+ i\epsilon^D J_{n+\muhat} \tilde{u} \right] \,
A_{n+\muhat}(\tilde{u},\cdots,\tilde{u},
{\hbox to 0pt{$u$}\raise-7pt
\hbox to -2pt{$\, {\scriptstyle \uparrow}$}
\raise-17pt \hbox to 9pt{${\scriptstyle -\muhat}$}},
\tilde{u}, \cdots, \tilde{u} \, ;J)
\biggl\}_{\displaystyle \tilde{u} \to u}
\nonumber
\\
&&=
\int [du] \, \, u \,
\biggl\{
\exp \biggl[ \,
-\frac{1}{2} \epsilon^{2-D}
\frac{\partial^2}{\partial \tilde{u}^2}
+\frac{1}{2} \epsilon^{2-D}
\frac{\partial^2}{\partial u^2}
\, \biggl] \,
\nonumber
\\
&&
{}~~~~~~~~~~
\times
\exp \biggl[ \,
- \epsilon^D V(u) + i\epsilon^D J_{n} u
\, \biggl]
\,
A_n(u,\cdots,u,
{\hbox to 0pt{$\tilde{u}$}\raise-7pt
\hbox to 0pt{$\, {\scriptstyle \uparrow}$}
\raise-17pt \hbox to 9pt{$\, {\scriptstyle \muhat}$}},
u, \cdots, u \, ;J)
\biggl\}_{\displaystyle \tilde{u} \to u}
\nonumber
\\
&&=
\int [du] \, \biggl(
u \,
\exp \left[ -\epsilon^D V(u) + i\epsilon^D J_n u \right]
\,
A_n(u, \cdots, u; J)
\nonumber
\\
&&
{}~~~~~
+ \epsilon^{2-D} \exp
\left[ -\epsilon^D V(u) + i\epsilon^D J_n u \right]
\biggl\{
\frac{\partial}{\partial \tilde{u}}
A_n(u,\cdots,u,
{\hbox to 0pt{$\tilde{u}$}\raise-7pt
\hbox to 0pt{$\, {\scriptstyle \uparrow}$}
\raise-17pt \hbox to 9pt{$\, {\scriptstyle \muhat}$}},
u, \cdots, u \, ;J)
\biggl\}_{\displaystyle \tilde{u} \to u}
\biggl) ,
\nonumber
\\
\eea
where in the last line we used the identity
\bea
\int dx \, x \,
\biggl\{
\exp \biggl[ \,
-\frac{1}{2} \epsilon^{2-D}
\frac{\partial^2}{\partial y^2}
+\frac{1}{2} \epsilon^{2-D}
\frac{\partial^2}{\partial x^2}
\, \biggl] \,
f(x,y)
\biggl\}_{\displaystyle y \to x}
{}~~~~~~~
\nonumber
\\
=
\int dx \,
\biggl(
x \, f(x,x) + \epsilon^{2-D}
\biggl\{
\frac{\partial}{\partial y} f(x,y)
\biggl\}_{\displaystyle y \to x}
\biggl)
\label{id2}
\eea
(See Appendix B for proof.)
Thus, we have shown the relation
\bea
&&
\int [du] \, u \, e^{-\epsilon^D V(u)} \, \epsilon^{-2}
\left[
A_{n+\muhat} (u,\cdots, u;J) \, e^{i\epsilon^D J_{n+\muhat}u}
- A_{n} (u,\cdots, u;J) \, e^{i\epsilon^D J_{n}u}
\right]
{}~~~
\nonumber
\\
&&=
\epsilon^{-D}
\int [du] \,
e^{-\epsilon^D V(u) + i\epsilon^D J_n u } \,
\biggl\{
\frac{\partial}{\partial \tilde{u}}
A_n(u,\cdots,u,
{\hbox to 0pt{$\tilde{u}$}\raise-7pt
\hbox to 0pt{$\, {\scriptstyle \uparrow}$}
\raise-17pt \hbox to 9pt{$\, {\scriptstyle \muhat}$}},
u, \cdots, u \, ;J)
\biggl\}_{\displaystyle \tilde{u} \to u} .
\eea
Taking sum over $\muhat$ in $2D$ directions, we find
\bea
&&
\int [du] \,
u \,
e^{-\epsilon^D V(u)}
\sum_{\muhat=1}^{2D} \epsilon^{-2}
\left[
A_{n+\muhat} (u,\cdots, u;J) \, e^{i\epsilon^D J_{n+\muhat}u}
- A_{n} (u,\cdots, u;J) \, e^{i\epsilon^D J_{n}u}
\right]
\nonumber
\\
&&=
\epsilon^{-D}
\int [du] \,
e^{-\epsilon^D V(u) + i\epsilon^D J_n u} \,
\frac{\partial}{\partial u}
A_n (u, \cdots, u;J) .
\eea
Then one may convert this equality
to the equation for the partition function $Z[J]$
using eqs.(\ref{condfora})
and (\ref{relzn}):
\bea
( \mbox{l.h.s.} ) &=&
\sum_{\muhat=1}^{2D} \epsilon^{-2}
\left\{
\frac{1}{i\epsilon^D}
\frac{\partial}{\partial J_{ \hbox to 7pt{$\scriptstyle n+\muhat$} } }
\, - \, \frac{1}{i\epsilon^D}
\frac{\partial}{\partial J_{n}}
\right\}
Z[J] ,
\rule[-8mm]{0mm}{6mm}
\\
( \mbox{r.h.s.} ) &=&
- \int [du] \,
\left[
- V'(u) + i J_n
\right] \,
e^{-\epsilon^D V(u) + i\epsilon^D J_n u}
\,
A_n (u, \cdots, u;J)
\rule[-4mm]{0mm}{6mm}
\nonumber
\\
&=&
\left\{
V' \biggl( \frac{1}{i\epsilon^D}
\frac{\partial}{\partial J_n} \biggl) - i J_n
\right\}
Z[J] .
\eea
Hence, the partition function satisfies
the Dyson-Schwinger equations (\ref{sdeq1}).
This shows that the Green functions constructed from the hole function
(\ref{greenfnfroma}) satisfy the Dyson-Schwinger equations
(\ref{sdeq2}).

\section{Extracting the Vacuum State at the Boundaries}
\cleqn

So far
we have discussed method for calculating Green functions that obey the
Dyson-Schwinger equations, while
we have left aside the issue
of their boundary conditions.
In fact, Green functions satisfy the Dyson-Schwinger equations
for arbitrary boundary conditions.
In quantum field theory, however,
we usually want to find the Green functions
given by the vacuum expectation values of time-ordered
field operator products.
We will see that such Green functions can be obtained by solving the
zeroth and first order local equations for the hole functions,
eqs.(\ref{zeroth}) and (\ref{first}), with appropriate
conditions.

The Hamiltonian $\hat{H}$
for the lattice field theory (\ref{pfn2}) is defined
from the transfer matrix of this theory:
\bea
e^{-\epsilon \, \hat{H}}
&=&
\exp \biggl[
- \frac{1}{2}\epsilon^D \sum_{l \in S}
\left\{
\sum_{\nuhat=1}^{D-1}
\frac{1}{2}
\biggl(
\frac{\phi_{l+\nuhat}-\phi_l}{\epsilon}
\biggl)^2
+ V(\phi_l)
\right\}
\biggl]
\nonumber
\\
&&\times
\exp \biggl[ \,
\frac{1}{2} \epsilon^{2-D} \sum_{l \in S}
\frac{\partial^2}{\partial \phi_l^2}
\, \biggl] \,
\exp \biggl[
- \frac{1}{2}\epsilon^D \sum_{l \in S}
\left\{
\sum_{\nuhat=1}^{D-1}
\frac{1}{2}
\biggl(
\frac{\phi_{l+\nuhat}-\phi_l}{\epsilon}
\biggl)^2
+ V(\phi_l)
\right\}
\biggl] ,
\label{timeevol}
\eea
where $\phi_l$'s are the fields on a
$(D \! - \! 1)$-dimensional lattice hyperplane $S$
corresponding to some fixed time.
$\hat{H}$ is hermitian,
and we denote the energy eigenstates as
\bea
\hat{H} \ket{\alpha} = E_\alpha \ket{\alpha} .
\eea
We can define Heisenberg operator $\hat{\Phi}_n$ just as
in the continuum theory.
Then the Green functions that obey the Dyson-Schwinger equations
(\ref{sdeq2}) is given by
\bea
\left< \phi_{i_1} \cdots \phi_{i_N} \right>
= \sum_{\alpha , \beta} C_{\alpha\beta}
\bra{\alpha} T \, \hat{\Phi}_{i_1} \cdots \hat{\Phi}_{i_N} \ket{\beta}
\label{boundarycond}
\eea
for arbitrary $C_{\alpha\beta}$'s.

Suppose we look for translationally invariant solutions
to the zeroth
order equations (\ref{zeroth}):
\bea
A_n^{(0)} = ~\mbox{independent of } n .
\label{tinva0}
\eea
Then any one-point function calculated from $A_n^{(0)}$ will also be
translationally invariant
\bea
\left< \phi_n^{\, m} \right>
=
\frac{
\rule[-2mm]{0mm}{6mm}
\int [du] \, u^m \,
A_n^{(0)}(u, \cdots, u)  \,
e^{-\epsilon^D V(u )}
}{
\rule[0mm]{0mm}{5mm}
\int [du] \,
A_n^{(0)} (u, \cdots, u) \,
e^{-\epsilon^D V(u )}
}
= \mbox{independent of } n .
\eea
It means we have selected a particular class of
boundary conditions in
(\ref{boundarycond}).
Since
\bea
\sum_{\alpha , \beta} C_{\alpha\beta}
\bra{\alpha} \hat{\Phi}_n^{\, m} \ket{\beta}
&=&
\sum_{\alpha , \beta} C_{\alpha\beta} \,
e^{-\tau (E_\beta -E_\alpha )}
\bra{\alpha} \hat{\Phi}_{(\vec{n},0)}^{\, m} \ket{\beta} ,
\eea
where we regard the $D$-th component of $n$ as the Euclidean time,
$n = (\vec{n},n_D)$ and $\tau = \epsilon \, n_D$,
we see the solutions (\ref{tinva0})
correspond to the boundary conditions
\bea
C_{\alpha\beta} = 0 ~~~\mbox{for}~E_\alpha \neq E_\beta .
\label{tinvboundcond}
\eea

Next, consider the solution $A_{n;k}^{(1)}$
to the first order equations
(\ref{first}), from which
one may calculate the two-point Green function
$\left< \phi_n \phi_k \right>$
subject to the same boundary conditions as that for the one-point
function, eq.(\ref{tinvboundcond}).
For $n_D > k_D$,
\bea
\left< \phi_n \phi_k \right>
&=&
\sum_{\alpha , \alpha'} C_{\alpha\alpha'}
\bra{\alpha} \hat{\Phi}_n \hat{\Phi}_k \ket{\alpha'}
{}~~~~~~~~~~
(E_\alpha=E_{\alpha'})
\nonumber
\\
&=&
\sum_{\alpha , \alpha' ,\beta} C_{\alpha\alpha'}
e^{-(\tau -\tau') (E_\beta -E_\alpha )}
\bra{\alpha} \hat{\Phi}_{(\vec{n},0)} \ket{\beta}
\bra{\beta} \hat{\Phi}_{(\vec{k},0)} \ket{\alpha'} ,
\eea
where $k = (\vec{k},k_D)$ and $\tau' = \epsilon \, k_D$.
Therefore, if we demand
the two-point function to be well-behaved as
$|n_D-k_D| \to \infty$, $E_\beta \geq E_\alpha$, so that
the vacuum state will be selected
at the boundaries.
It is the condition that should be imposed on $A_{n;k}^{(1)}$.

To calculate higher Green functions satisfying the right boundary
conditions, solve successively higher order local equations for higher
order hole functions using thus obtained $A_n^{(0)}$ and
$A_{n;k}^{(1)}$.

\section{Basic Properties of the Local Equations}
\cleqn

\subsection{$D=1$ Case}

Let us examine the local equation (\ref{localeq})
for the 1-dimensional lattice with $N+1$ sites
($n=0,1, \ldots, N$)
:
\bea
e^{\frac{1}{2}\epsilon \frac{\partial^2}{\partial u^2}} \,
e^{-\epsilon V(u)+i\epsilon J_n u} \,
A_n (u, \tilde{u};J)
=
e^{\frac{1}{2}\epsilon \frac{\partial^2}{\partial \tilde{u}^2}} \,
e^{-\epsilon V(\tilde{u})+i\epsilon J_{n+1} \tilde{u}} \,
A_{n+1} (u, \tilde{u};J) .
\label{dim1eq}
\eea
Noting that $A_n (u, \tilde{u};J)$ is independent of $J_n$,
the solution is found to be
\bea
A_n (u, \tilde{u};J) &=&
\sum_i
\left[
e^{\frac{1}{2}\epsilon \frac{\partial^2}{\partial u^2}} \,
e^{-\epsilon V(u)+i\epsilon J_{n-1} u} \,
\cdots
e^{\frac{1}{2}\epsilon \frac{\partial^2}{\partial u^2}} \,
e^{-\epsilon V(u)+i\epsilon J_{0} u} \,
f_i (u)
\right.
\nonumber
\\
&&~~~
\times
\left.
e^{-\epsilon V(\tilde{u})+i\epsilon J_{n+1} \tilde{u}} \,
e^{\frac{1}{2}\epsilon \frac{\partial^2}{\partial \tilde{u}^2}} \,
\cdots
e^{-\epsilon V(\tilde{u})+i\epsilon J_{N} \tilde{u}} \,
e^{\frac{1}{2}\epsilon \frac{\partial^2}{\partial \tilde{u}^2}} \,
g_i (\tilde{u})
\right]
\label{dim1sol}
\eea
for $\forall f_i (u), ~g_i (\tilde{u})$.

Now we take the continuum limit $\epsilon \to 0$
fixing $T = \epsilon \, N$ and
$\tau = \epsilon \, n$.
Eq.(\ref{dim1eq}) reduces to
\bea
\frac{\partial}{\partial \tau}
A (\tau ; u, \tilde{u}; J)
=
\left[
\left( \hat{H}_{\tilde{u}}
- i J(\tau) \tilde{u}
\right)
-
\left( \hat{H}_u
- i J(\tau) u
\right)
\right]
A (\tau ; u, \tilde{u}; J) ,
\label{continuum1}
\eea
where
\bea
\hat{H}_u &=&
-\frac{1}{2} \frac{\partial^2}{\partial u^2}
+ V(u) , ~~~~~
\hat{H}_{\tilde{u}} =
-\frac{1}{2} \frac{\partial^2}{\partial \tilde{u}^2}
+ V(\tilde{u}) ,
\label{continuum2}
\eea
and the hole function (\ref{dim1sol})
to
\bea
A (\tau ; u, \tilde{u}; J) =
\sum_i
\bra{g_i}
T \, e^{-\int^T_\tau d\tau' \, \left(
\hat{H} - i J(\tau') \hat{\phi} \right)}
\ket{\tilde{u}}
\bra{u}
T \, e^{-\int^\tau_0 d\tau' \, \left(
\hat{H} - i J(\tau') \hat{\phi} \right)}
\ket{f_i} ,
\label{continuum3}
\eea
which are the slightly generalized forms of the
equations discussed in
Section 1.
Then one obtains the correct partition function of the theory
from the hole function
\bea
Z[J(\tau)] &=& \lim_{\epsilon \to 0} \,
\int du \, \, e^{-\epsilon V(u) + i\epsilon J_n u}  \,
A_n (u,u;J)
\\
&=&
\sum_i
\bra{g_i}
T \, e^{-\int^T_0 d\tau' \, \left(
\hat{H} - i J(\tau') \hat{\phi} \right)}
\ket{f_i} .
\eea
The Green functions are obtained from $Z[J]$ (or from
$A (\tau ; u, \tilde{u}; J)$), and it is
easy to check that the procedure indicated in the previous section
picks up the vacuum $\ket{f_i}, \, \ket{g_i} \to \ket{0}$.

\subsection{Formal Solutions}

We present formal solutions to the local equations (\ref{localeq}).

First consider the hole function (\ref{defholefn}) in the case of free
field theory $V(\phi)= \frac{1}{2} m^2 \phi^2$.
Since the integrand is Gaussian, one can explicitly perform
the integration.
We find
\bea
A_n^{(free)} (u_1, \cdots, u_{2D} ; J) =
\exp \left[
-\frac{1}{2} \left( \vec{u} + \vec{J} \right)^T
N^{-1} \left( \vec{u} + \vec{J} \right) -
\frac{1}{2} \epsilon^{D-2}
\sum_{\muhat =1}^{2D} u_{\muhat}^2
\right] ,
\label{freeholefn}
\eea
where
\bea
\begin{array}{ccl}
\left( \vec{J} \right)_l &=& \epsilon^D J_l ,
\rule[-4mm]{0mm}{6mm}
\\
\left( \vec{u} \right)_l &=&
{\displaystyle
\sum_{\muhat=1}^{2D}
\epsilon^{D-2} \, u_{\muhat} \, \delta_{n+\muhat ,l}
} ,
\rule[-3mm]{0mm}{6mm}
\\
\left( N^{-1} \right)_{lk} &=&
G(l-k) -
{\displaystyle
\frac{1}{G(0)}
} \,
G(l-n) G(n-k) ,
\end{array}
\eea
and
\bea
G(l) = \frac{1}{\epsilon^D}
\int_B
{\textstyle
\frac{d^Dp}{(2\pi )^D}
} \,
\frac{e^{ip \cdot l}}
{m^2 +
2 \epsilon^{-2} \sum_{\muhat=1}^{D} [ 1 - \cos p_\mu  ]}
,
{}~~~~~~~
B = (-\pi , \pi )^D
{}.
\eea
It is straightforward to verify
that eq.(\ref{freeholefn}) satisfies the
local equations (\ref{localeq}).

For a general potential $V(\phi )$, we write
\bea
V(\phi ) = \frac{1}{2} m^2 \phi^2 + {\cal L}_{int}(\phi ) .
\eea
Then from the definition (\ref{defholefn}),
the formal solution can be written as
\bea
A_n (u_1, \cdots, u_{2D} ; J) =
\exp \biggl[
- \epsilon^D \sum_{l \in \Lambda} {\cal L}_{int}
\biggl( \frac{1}{i\epsilon^D} \frac{\partial}{\partial J_l}
\biggl)
\biggl] \, \,
A_n^{(free)} (u_1, \cdots, u_{2D} ; J) .
\eea
Again, it is straightforward to show
that this formal solution satisfies the local
equations (\ref{localeq}).

\subsection{Reduction of Variables}

The closed equations (\ref{zeroth}) and (\ref{first})
are equations for the hole functions of $2D$ link variables.
We shall turn
the problem of solving these equations
into that of solving coupled
linear equations for a set of functions of
{\it two} variables.
When the system in question is translationally invariant as well as
invariant under rotation by $90^{\circ}$, we define
\bea
&&
A(u,\tilde{u}) \equiv
A_n^{(0)}(u,\cdots,u,
{\hbox to 0pt{$\tilde{u}$}\raise-7pt
\hbox to 0pt{$\, {\scriptstyle \uparrow}$}
\raise-17pt \hbox to 9pt{$\, {\scriptstyle \muhat}$}},
u, \cdots, u )
{}~:~~~
\mbox{independent of}~n, \muhat ,
\label{redv0}
\\&&
B_{n-k;\muhat}(u,\tilde{u}) \equiv
A_{n;k}^{(1)}(u,\cdots,u,
{\hbox to 0pt{$\tilde{u}$}\raise-7pt
\hbox to 0pt{$\, {\scriptstyle \uparrow}$}
\raise-17pt \hbox to 9pt{$\, {\scriptstyle \muhat}$}},
u, \cdots, u ) .
\label{redv1}
\eea
Then eqs.(\ref{zeroth})-(\ref{cond})
are rewritten as
\bea
&&
e^{\frac{1}{2} \epsilon^{2-D} \frac{\partial^2}{\partial u^2}} \,
e^{-\epsilon^D V(u)} \,
A(u,\tilde{u}) =
e^{\frac{1}{2} \epsilon^{2-D}
\frac{\partial^2}{\partial \tilde{u}^2}} \,
e^{-\epsilon^D V(\tilde{u})} \,
A(\tilde{u},u) ,
\rule[-4mm]{0mm}{6mm}
\label{redv2}
\\
&&
e^{\frac{1}{2} \epsilon^{2-D} \frac{\partial^2}{\partial u^2}} \,
e^{-\epsilon^D V(u)} \,
B_{l;\muhat}(u,\tilde{u}) -
e^{\frac{1}{2} \epsilon^{2-D}
\frac{\partial^2}{\partial \tilde{u}^2}} \,
e^{-\epsilon^D V(\tilde{u})} \,
B_{l+\muhat;-\muhat}(\tilde{u},u)
\nonumber
\\
&& ~~~~~
=
\delta_{l+\muhat,0} \,
e^{\frac{1}{2} \epsilon^{2-D}
\frac{\partial^2}{\partial \tilde{u}^2}} \,
e^{-\epsilon^D V(\tilde{u})} \,
\tilde{u} \, A(\tilde{u},u)
- \delta_{l,0} \,
e^{\frac{1}{2} \epsilon^{2-D} \frac{\partial^2}{\partial u^2}} \,
e^{-\epsilon^D V(u)} \,
u \, A(u,\tilde{u}) ,
\rule[-4mm]{0mm}{6mm}
\label{redv3}
\\
&&
B_{0;\muhat}(u,\tilde{u}) = 0.
\eea
In addition, the
following conditions should be satisfied
according to the definitions (\ref{redv0}) and (\ref{redv1}).
\bea
&&
\rule[-7mm]{0mm}{6mm}
\frac{d}{du} A(u,u) =
2D \left[
\frac{\partial}{\partial \tilde{u}}
A(u,\tilde{u})
\right]_{\displaystyle \tilde{u} \to u} ,
\label{addcond1}
\\
&&
\rule[-4mm]{0mm}{6mm}
B_{l;\muhat}(u,u) = B_l(u) ~:~~~
\mbox{independent of}~\muhat ,
\\
&&
\frac{d}{du} \, B_l(u)
=
\sum_{\muhat=1}^{2D}
\left[
\frac{\partial}{\partial \tilde{u}}
B_{l;\muhat}(u,\tilde{u})
\right]_{\displaystyle \tilde{u} \to u} .
\label{addcond2}
\eea
Following the discussion in Section 4,
one can verify that when
$A(u,\tilde{u})$ and
$B_{l;\muhat}(u,\tilde{u})$ satisfy
eqs.(\ref{redv2})-(\ref{addcond2}),
the one-point and two-point Green functions constructed from these hole
functions obey the Dyson-Schwinger equations (\ref{sdeq2}).
Thus, it suffices to solve these set of coupled linear equations for
$A(u,\tilde{u})$ and
$B_{l;\muhat}(u,\tilde{u})$ to obtain the one-point and two-point
Green functions.

\section{Summary and Discussion}

In Section 2 we review the Dyson-Schwinger equations for a Euclidean
lattice scalar field theory.
We argue that they carry full
information of the theory
since the equations are
identified with the Fourier transform of the defining equations for
the weight factor $e^{-S[\phi]}$.

In Section 3 we define the ``hole function'', $A_n$, associated with a
lattice site $n$.
It is defined such that any $N$-point Green function can be constructed
from it.
We see that the hole function obeys a set of local equations,
which are shown to be equivalent to the Dyson-Schwinger
equations (Section 4).
The remarkable feature is that
the local equations, when expanded in terms of
source $J$,
reduce to sets of {\it closed} equations for the hole functions.
It is in contrast to the
Dyson-Schwinger equations which are the infinite series of coupled
equations.

To obtain the Green functions satisfying the right boundary conditions:
Find translationally invariant solutions $A^{(0)}_n$ to the zeroth order
equation (\ref{zeroth}).
Then solve the first order equation (\ref{first}) for $A_{n;k}^{(1)}$,
and demand that the two-point function calculated from $A_{n;k}^{(1)}$
to behave well as $|n-k|\to \infty$
(Section 5).

Section 6 summarizes
some basic properties of the local equations.
We have seen
that the equation correctly reproduces the known results from
quantum mechanics when $D=1$ (Subsection 6.a).
The existence of
formal solution to the local equations is anticipated from the
ordinary perturbation theory, and the explicit form is given
(Subsection 6.b).
The closed sets of local equations for the hole functions can be
reduced to the
coupled linear equations for functions of two variables.
(Subsection 6.c).

\medbreak

It would be straightforward to solve numerically
the closed local equations given in the form of Subsection 6.c.
If we expand the functions $A(u,\tilde{u})$ and
$B_{l;\muhat}(u,\tilde{u})$ in terms of some appropriate
functional bases, task to solve
the equations would reduce to matrix calculations.

Finally, it may be instructive to see
which part of the information on the
Dyson-Schwinger equations is contained in
each set of closed equations.
Consider the solution $A_n^{(0)}$
to the simplest equations (\ref{zeroth}).
Since we may obtain
$F_{n,\muhat}|_{J=0}$
from $A_n^{(0)}$ via eq.(\ref{relaandf3}),
not only the one-point functions $\left< \phi_n^{\, m}
\right>$ but also two-point functions
of the nearest neighbors
$\left< \phi_n^{\, m} \phi_{n+\muhat}^{\, m'} \right>$
can be constructed from $A_n^{(0)}$.\footnote{
Since we defined the hole function to be a function of links surrounding
$n$ rather than of the field on $n$, it carries some information on the
nearest neighbors.
}
In fact, one can show using the zeroth order equations (\ref{zeroth})
alone
the following particular part of the Dyson-Schwinger equations
\bea
\left< ( \dal_n \phi_n ) \, \phi_n^{\, m}
- V'(\phi_n) \, \phi_n^{\, m} + \frac{m}{\epsilon^D} \,
\phi_n^{\, m-1} \right> = 0 ,
{}~~~~~
(m = 0, 1, 2, \ldots ) .
\eea
Similarly, from the solutions to eqs.(\ref{zeroth}) and (\ref{first}),
one can show
\bea
\left< ( \dal_n \phi_n ) \, \phi_n^{\, m} \phi_k
- V'(\phi_n) \, \phi_n^{\, m} \phi_k
+ \frac{m}{\epsilon^D} \, \phi_n^{\, m-1} \phi_k
+ \frac{1}{\epsilon^D} \, \delta_{n,k} \, \phi_n^{\, m}
\right> = 0 ,
{}~~~~~
\nonumber
\\
(m = 0, 1, 2, \ldots ) .
\eea

\medbreak

The author is grateful to K. Hikasa for carefully reading the manuscript
and making useful comment.

\section*{Appendix A: Minkowski Versions}
\renewcommand{\theequation}{A.\arabic{equation}}
\cleqn
\clfn

We present here the local equations for Minkowski space-time.

Define
\bea
\eta_{\muhat} = \left\{
\begin{array}{cll}
-1 & \mbox{for}~
\muhat = \pm \hat{x}_1 , \ldots , \pm \hat{x}_{D-1}
& \mbox{(space direction)} \\
+1 & \mbox{for}~ \muhat = \pm \hat{x}_D &
\mbox{(time direction)}
\end{array}
\right. .
\eea
The action is given by
\bea
S[\phi ] =
\sum_{l} \epsilon^D
\biggl\{ \,
\sum_{\muhat =1}^D \frac{1}{2} \eta_{\muhat}
\biggl( \frac{\phi_{l+\muhat} - \phi_l}{\epsilon} \biggl)^2
- V(\phi_l ) \, \biggl\} ,
\eea
and the hole function is defined by
\bea
&&
A_n (u_1, \cdots , u_{2D}; J)
\nonumber
\\
&& \equiv
\int \prod_{l \neq n} [ d\phi_l ] \,
\exp \biggl[
i S_n'[\phi]
+ i\epsilon^D \sum_{l \neq n}J_l \phi_l
+ i \epsilon^D \sum_{\muhat = 1}^{2D} \frac{1}{2}
\eta_{\muhat}
\biggl( \frac{\phi_{n+\muhat}-u_{\muhat}}{\epsilon} \biggl)^2
\biggl]
\label{defmholefn}
\eea
with
\bea
S_n'[\phi] = S[\phi] - \epsilon^D
\biggl\{ \,
\sum_{\muhat =1}^{2D} \frac{1}{2} \eta_{\muhat}
\biggl( \frac{\phi_{n+\muhat} - \phi_n}{\epsilon} \biggl)^2
- V(\phi_l ) \, \biggl\} .
\eea
Here, the integral measure is defined as
$[d\phi_l] = d\phi_l / \sqrt{2\pi i\epsilon}$.

The local equations satisfied by the hole function is
\bea
&&
\exp \biggl[ \,
\frac{1}{2} i \epsilon^{2-D} \eta_{\muhat}
\frac{\partial^2}{\partial u^2}
\, \biggl] \,
\exp \biggl[ \,
- i \epsilon^D V(u) + i\epsilon^D J_{n} u
\, \biggl]
\,
A_n(u,\cdots,u,
{\hbox to 0pt{$\tilde{u}$}\raise-7pt
\hbox to 0pt{$\, {\scriptstyle \uparrow}$}
\raise-17pt \hbox to 9pt{$\, {\scriptstyle \muhat}$}},
u, \cdots, u \, ;J)
\nonumber
\\
&&
=
\exp \biggl[ \,
\frac{1}{2} i \epsilon^{2-D} \eta_{\muhat}
\frac{\partial^2}{\partial \tilde{u}^2}
\, \biggl] \,
\exp \biggl[ \,
- i \epsilon^D V(\tilde{u}) + i\epsilon^D J_{n+\muhat} \tilde{u}
\, \biggl]
\,
A_{n+\muhat}(\tilde{u},\cdots,\tilde{u},
{\hbox to 0pt{$u$}\raise-7pt
\hbox to -2pt{$\, {\scriptstyle \uparrow}$}
\raise-17pt \hbox to 9pt{${\scriptstyle -\muhat}$}},
\tilde{u}, \cdots, \tilde{u} \, ;J) ,
\nonumber
\\
\label{minkowski1}
\eea
where a condition that follows from the definition eq.(\ref{defmholefn})
is
\bea
A_n (u_1, \cdots , u_{2D}; J)
= \, \mbox{independent of}~J_n .
\eea
An identity corresponding to eq.(\ref{difint}) is
given by
\bea
\int \frac{dy}{\sqrt{2\pi i a}} \,
e^{i(x-y)^2/2a} \, f(y) =
e^{\frac{1}{2}i a\frac{d^2}{dx^2}} \, f(x) .
\eea

When we expand the local equation (\ref{minkowski1}) in $J$, the
zeroth-order and first-order equations are, respectively,
\bea
&&
\exp \biggl[ \,
\frac{1}{2} i \epsilon^{2-D} \eta_{\muhat}
\frac{\partial^2}{\partial u^2}
\, \biggl] \,
\exp \biggl[ \,
- i \epsilon^D V(u)
\, \biggl]
\,
A_n^{(0)}(u,\cdots,u,
{\hbox to 0pt{$\tilde{u}$}\raise-7pt
\hbox to 0pt{$\, {\scriptstyle \uparrow}$}
\raise-17pt \hbox to 9pt{$\, {\scriptstyle \muhat}$}},
u, \cdots, u )
\nonumber
\\
&&=
\exp \biggl[ \,
\frac{1}{2} i \epsilon^{2-D} \eta_{\muhat}
\frac{\partial^2}{\partial \tilde{u}^2}
\, \biggl] \,
\exp \biggl[ \,
-i \epsilon^D V(\tilde{u})
\, \biggl]
\,
A_{n+\muhat}^{(0)}(\tilde{u},\cdots,\tilde{u},
{\hbox to 0pt{$u$}\raise-7pt
\hbox to -2pt{$\, {\scriptstyle \uparrow}$}
\raise-17pt \hbox to 9pt{${\scriptstyle -\muhat}$}},
\tilde{u}, \cdots, \tilde{u} )
\eea
and
\bea
&&
\exp \biggl[ \,
\frac{1}{2} i \epsilon^{2-D} \eta_{\muhat}
\frac{\partial^2}{\partial u^2}
\, \biggl] \,
\exp \biggl[ \,
- i \epsilon^D V(u)
\, \biggl]
\,
A_{n;k}^{(1)}(u,\cdots,u,
{\hbox to 0pt{$\tilde{u}$}\raise-7pt
\hbox to 0pt{$\, {\scriptstyle \uparrow}$}
\raise-17pt \hbox to 9pt{$\, {\scriptstyle \muhat}$}},
u, \cdots, u )
\nonumber
\\
&&
-
\exp \biggl[ \,
\frac{1}{2} i \epsilon^{2-D} \eta_{\muhat}
\frac{\partial^2}{\partial \tilde{u}^2}
\, \biggl] \,
\exp \biggl[ \,
- i \epsilon^D V(\tilde{u})
\, \biggl]
\,
A_{n+\muhat ;k}^{(1)}(\tilde{u},\cdots,\tilde{u},
{\hbox to 0pt{$u$}\raise-7pt
\hbox to -2pt{$\, {\scriptstyle \uparrow}$}
\raise-17pt \hbox to 9pt{${\scriptstyle -\muhat}$}},
\tilde{u}, \cdots, \tilde{u} )
\nonumber \\
&=&
\delta_{n+\muhat,k} \,
\exp \biggl[ \,
\frac{1}{2} i \epsilon^{2-D} \eta_{\muhat}
\frac{\partial^2}{\partial \tilde{u}^2}
\, \biggl] \,
\exp \biggl[ \,
- i\epsilon^D V(\tilde{u})
\, \biggl]
\, \tilde{u} \,
A_{n+\muhat}^{(0)}(\tilde{u},\cdots,\tilde{u},
{\hbox to 0pt{$u$}\raise-7pt
\hbox to -2pt{$\, {\scriptstyle \uparrow}$}
\raise-17pt \hbox to 9pt{${\scriptstyle -\muhat}$}},
\tilde{u}, \cdots, \tilde{u} )
\nonumber \\
&& -
\delta_{n,k} \,
\exp \biggl[ \,
\frac{1}{2} i \epsilon^{2-D} \eta_{\muhat}
\frac{\partial^2}{\partial u^2}
\, \biggl] \,
\exp \biggl[ \,
- i \epsilon^D V(u)
\, \biggl]
\, u \,
A_n^{(0)}(u,\cdots,u,
{\hbox to 0pt{$\tilde{u}$}\raise-7pt
\hbox to 0pt{$\, {\scriptstyle \uparrow}$}
\raise-17pt \hbox to 9pt{$\, {\scriptstyle \muhat}$}},
u, \cdots, u )  ,
\eea
where
\bea
A_n^{(0)} & \equiv & A_n \biggl|_{J=0} ,
\rule[-4mm]{0mm}{6mm}
\\
A_{n;k}^{(1)} & \equiv &
\frac{1}{i\epsilon^D} \,
\frac{\partial}{\partial J_k} \, A_n \biggl|_{J=0} ,
\eea
and
\bea
A_{n;n}^{(1)} (u_1,\cdots,u_{2D}) = 0 .
\eea

\section*{Appendix B}
\renewcommand{\theequation}{B.\arabic{equation}}
\cleqn

In this appendix, we prove eqs.(\ref{id1}) and (\ref{id2}):
\bea
&&
\int dx \,
\biggl\{
\exp \biggl[ \,
-\frac{1}{2} \epsilon^{2-D}
\frac{\partial^2}{\partial y^2}
+\frac{1}{2} \epsilon^{2-D}
\frac{\partial^2}{\partial x^2}
\, \biggl] \,
f(x,y)
\biggl\}_{\displaystyle y \to x}
=
\int dx \, f(x,x) ,
\eea
and
\bea
\int dx \, x \,
\biggl\{
\exp \biggl[ \,
-\frac{1}{2} \epsilon^{2-D}
\frac{\partial^2}{\partial y^2}
+\frac{1}{2} \epsilon^{2-D}
\frac{\partial^2}{\partial x^2}
\, \biggl] \,
f(x,y)
\biggl\}_{\displaystyle y \to x}
{}~~~~~~~
\nonumber
\\
=
\int dx \,
\biggl(
x \, f(x,x) + \epsilon^{2-D}
\biggl\{
\frac{\partial}{\partial y} f(x,y)
\biggl\}_{\displaystyle y \to x}
\biggl) .
\eea

Here, we assume that the function $f(x,x)$ swiftly vanishes
as $|x| \to \infty$, as well as\\
$
e^{
-\frac{1}{2} \epsilon^{2-D}
\frac{\partial^2}{\partial y^2}
+\frac{1}{2} \epsilon^{2-D}
\frac{\partial^2}{\partial x^2}
}
f(x,y)
$
swiftly vanishes as $|x|, |y| \to \infty$.

We write $f(x,y) = \bra{x} \hat{f} \ket{y}$.
Then, we see
\bea
&&
\int dx \,
\biggl\{
\exp \biggl[ \,
-\frac{1}{2} \epsilon^{2-D}
\frac{\partial^2}{\partial y^2}
+\frac{1}{2} \epsilon^{2-D}
\frac{\partial^2}{\partial x^2}
\, \biggl] \,
f(x,y)
\biggl\}_{\displaystyle y \to x}
\nonumber
\\
&&=
\int dx \, \bra{x} e^{-\frac{1}{2} \epsilon^{2-D} \hat{p}^2} \,
\hat{f} \,
e^{\frac{1}{2} \epsilon^{2-D} \hat{p}^2 }\ket{x}
\nonumber
\\
&&=
\mbox{Tr} \left[
e^{-\frac{1}{2} \epsilon^{2-D} \hat{p}^2} \,
\hat{f} \,
e^{\frac{1}{2} \epsilon^{2-D} \hat{p}^2}
\right]
= \mbox{Tr} \left[ \hat{f} \right]
\nonumber
\\
&& =
\int dx \, f(x,x) .
\eea
Also,
\bea
&&
\int dx \, x \,
\biggl\{
\exp \biggl[ \,
-\frac{1}{2} \epsilon^{2-D}
\frac{\partial^2}{\partial y^2}
+\frac{1}{2} \epsilon^{2-D}
\frac{\partial^2}{\partial x^2}
\, \biggl] \,
f(x,y)
\biggl\}_{\displaystyle y \to x}
{}~~~~~~~
\nonumber
\\
&&=
\int dx \, \bra{x}
e^{-\frac{1}{2} \epsilon^{2-D} \hat{p}^2} \,
\hat{f} \,
e^{\frac{1}{2} \epsilon^{2-D} \hat{p}^2} \, \hat{x}
\ket{x}
\nonumber
\\
&&=
\mbox{Tr} \left[
e^{-\frac{1}{2} \epsilon^{2-D} \hat{p}^2} \,
\hat{f} \,
e^{\frac{1}{2} \epsilon^{2-D} \hat{p}^2} \, \hat{x}
\right]
\nonumber
\\
&&=
\mbox{Tr} \left[
\hat{f} \,
\left(
e^{\frac{1}{2} \epsilon^{2-D} \hat{p}^2} \,
\hat{x} \,
e^{-\frac{1}{2} \epsilon^{2-D} \hat{p}^2}
\right)
\right]
=
\mbox{Tr} \left[
\hat{f} \,
\left(
\hat{x} - i\epsilon^{2-D} \hat{p}
\right)
\right]
\nonumber
\\
&&=
\int dx \,
\biggl(
x \, f(x,x) + \epsilon^{2-D}
\biggl\{
\frac{\partial}{\partial y} f(x,y)
\biggl\}_{\displaystyle y \to x}
\biggl) .
\eea



\newpage

\section*{Figure Captions}
\renewcommand{\labelenumi} {Fig. \arabic{enumi}}
\begin{enumerate}

\item 

The hole function $A_n(u_1,\cdots,u_{2D};J)$ is defined as a function of
link variables surrounding the site $n$.

\item 

The function $F_{n,\muhat}$ is defined as a function of link variables
surrounding two adjacent sites $n$ and $n+\muhat$;
$n$ is surrounded by $u_1,\ldots,u_{2D}$ (except $u_{\muhat}$), and
$n+\muhat$ is surrounded by $\tilde{u}_1,\ldots,\tilde{u}_{2D}$
(except $\tilde{u}_{-\muhat}$).
Note that the link connecting $n$ and $n+\muhat$ is missing.

\end{enumerate}

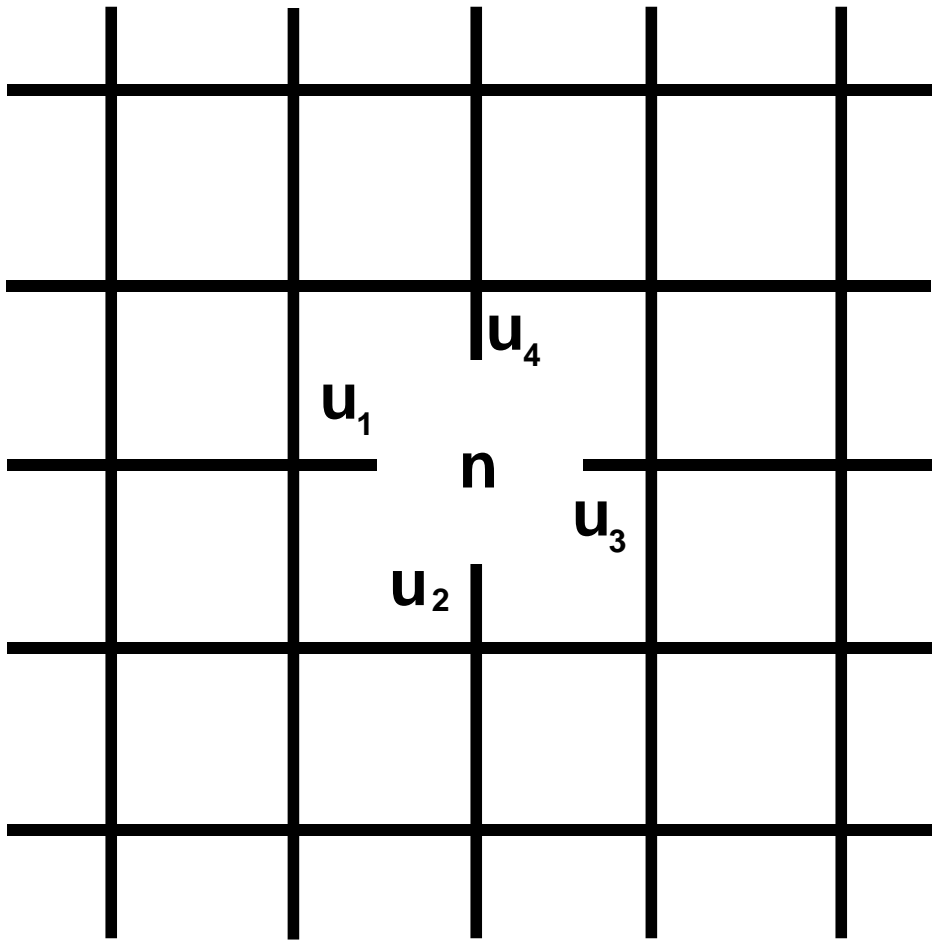
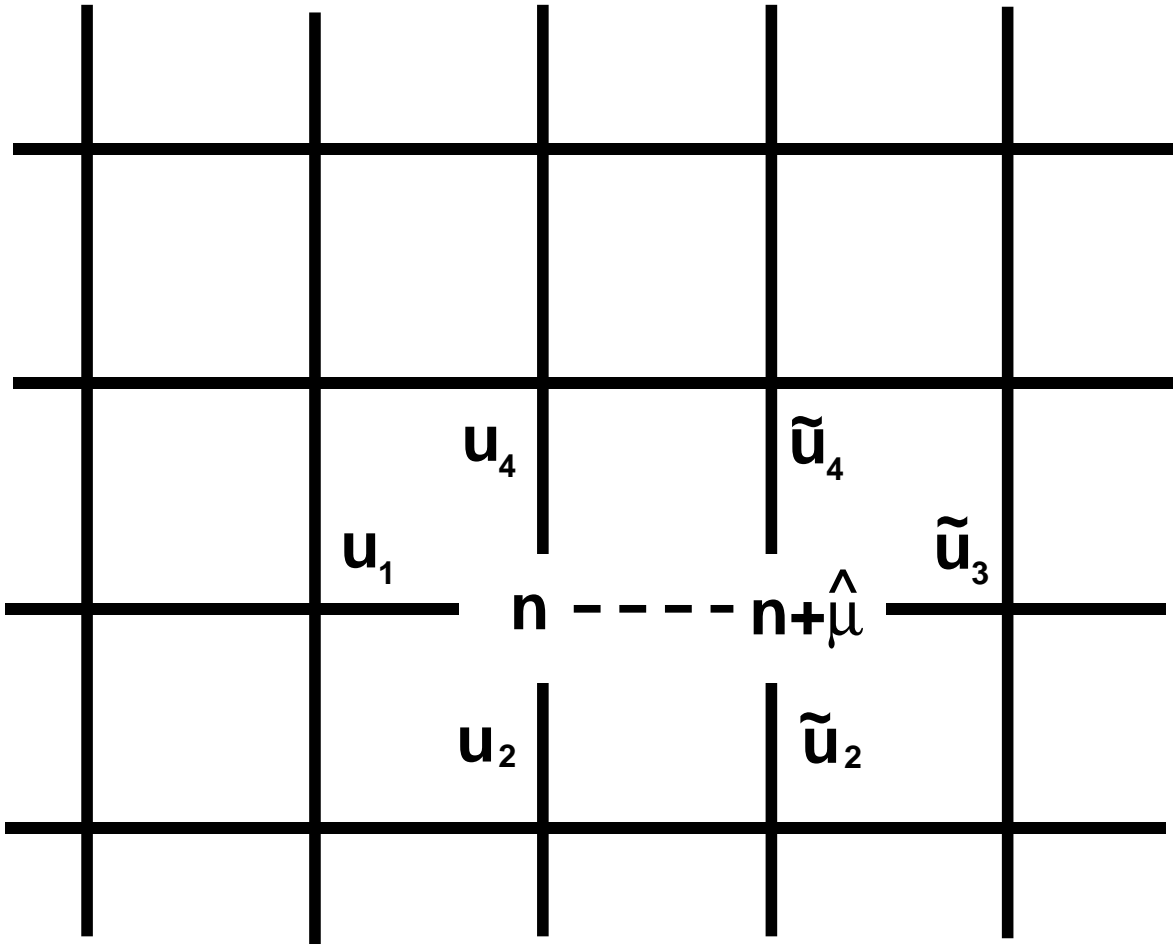
\end{document}